# Mathematical definition of public language, and modeling of will and consciousness based on the public language

Hana Hebishima, Mina Arakaki, Chikako Dozono, Hanna Frolova, Shinichi Inage

To propose a mathematical model of consciousness and will, we first simulated the inverted qualia with a toy model of a neural network. As a result, we confirmed that there can be an inverted qualia on the neural network. In other words, the qualia were individual-dependent and considered difficult as an indicator of consciousness and will. To solve that difficulty, we introduce a probability space and a random variable into a set of qualia and define a public language for events. Based on this idea of public language, consciousness and will are modeled. In this proposal, future actions are randomly selected from the comparison between "recognition of events" by external observation and past episodic memory, and the actual "recognition of actions" is regarded as the occurrence of consciousness. The basic formula is also derived. This proposal is compared with other past philosophical discussions.

*Key words: conscious, will, mathematical model, probability space, information entropy*

## 1. Introduction

This paper is concerned with the mathematical definition of public language and the model of consciousness derived from it. Modeling of consciousness is an important topic, and the relationship between the conscious realm and its physical realm has been studied and developed over the centuries, involving not only scientists but also philosophers and theologians (Bill Faw, 2014). Examples of models of consciousness include the Global Workspace Theory (Stanislas Dehaene et al., 1998), Multiple Draft Theory (Daniel C Dennett, 1993), Higher Thought Theory (Peter Carruthers, 2016), Dehaene‐Changeux model (Dehaene S, Changeux JP, 2000), Integrated Information Theory (Tononi G, 2004, Masafumi Oizumi aet al., 2014) and many others. There are also many good reviews of those models (Leonid I. Perlovsky, 2006, Christopher Durugbo et al., 2013), and we won't be introducing each model sequentially here.

Among those theories, the integrated information theory was proposed by Tononi to quantify consciousness‐ especially using the concept of information in information theory. The main purpose of the program is to quantitatively evaluate the perception of a person as a first-person. Integrated information theory does not keep in mind the so-called "hard problem of consciousness" but starts from "consciousness exists" and adopts "information," "integration," "structure" and "exclusion" as axioms. This system of theories is also of great interest, especially in terms of quantifying the functions that complex networks produce. The basic concept is that consciousness appears when much information is integrated, and the degree of integration is expressed by the integrated information quantity:Φ. The Φ is defined for a set of variables that interact and evolve over time, and is a quantification of how much more information the entire original network produces compared to the sum of the information produced by the subnetworks that divided the network. For quantification, we use the concept of Kullback–Leibler divergence, which is also used in this paper.

These many models of consciousness complement metaphysical theories of consciousness, such as functionalism, identity theory, dialogic dualism, and neutral monism. The following is a summary of the topics related to this paper.

### 1.1 Qualia

"A unique texture that cannot be explained by words felt by each individual, caused by subjective experiences and senses" is qualia. Even if we express our thoughts and feelings in words, they are essentially things only we can know. For example, let's say a person gets a "bad vibe" when they see a particular color. It is impossible to communicate and share that specific "bad feeling" with others by any means. And a hideous color for one person may be a favorable color for another. However, while we can explain why it is preferable, we cannot verbalize how it is preferable. In other words, because qualia refer to the very feeling of everyone, it is impossible for others to feel qualia generated within them in the same way.

Qualia is told with the following two characteristics.
1) Qualia is a subjective feeling gained through individual experience and the difficulty of verbalizing everything correctly
2) Qualia are extremely personal and subjective and cannot be shared with others, and even if we go through the same experience, the qualia we gain is incomprehensible to others

Thus, all sensations take place in an individual's brain and do not achieve the exact same qualia as others.

### 1.2 Philosophical Zombies

The "philosophical zombie" is a thought experiment designed to counter the physicalism of qualia and perceptions of consciousness. used by Chalmers, D., to describe qualia (Chalmers, D., 1996). A philosophical zombie is a thought experiment defined as "a being who behaves like a normal human being but in fact has no inner feelings." A major characteristic of philosophical zombies is the lack of qualia. For a philosophical zombie, all emotions and sensations are just part of a brain activity. To put it simply, a philosophical zombie is an entity that has sensations and emotions as a function but no sensations and emotions as a reality. Philosophical zombies are said to be unrecognizable because they look and behave no differently from humans. This is because, as mentioned above, qualia refer to "the sensations and emotions felt by an individual," and it is impossible to prove whether the other person has qualia. If it exists but can't be recognized, no one can prove it.

Physicism for consciousness is a monism that ascribes all phenomena to physics. The position is that the human mind and senses fluctuate due to some physical phenomenon and can be observed as an activity of the brain, and for this reason, in physicalism, the human mind and qualia are also regarded as "physical objects."

An opposing position is the dualistic idea that there are objects that can be physically observed and objects that cannot be physically observed. Even if we try to physically analyze emotions and sensations, what we can observe is the physical phenomena of the brain, and we can't observe whether there is "something" other than the physical phenomena of the brain. This leads to the idea that the essence of the human mind and consciousness cannot be expressed in modern physics, and this is called the "hard problem of consciousness."

In physicalism, the "mind," such as consciousness and qualia, is taken as a physical phenomenon of the brain - that is, "the mind is the same as the senses and emotions as functions." If physicalism is right, we can say that philosophical zombies are unimaginable and "improbable." Physicism, however, cannot prove why philosophical zombies are improbable, because the physical phenomena of the brain do not explain why or how the mind arises. Based on this logic, the argument is that "Physicism is wrong if philosophical zombies are imaginable, and their existence cannot be denied."

### 1.3 Inverted qualia

Consider, for example, the color red. Observers watching the same sunset: Even if A and B each recognize the color of the sunset as "red," they cannot confirm whether the private "red" that A and B recognize is the same. For example, even if A's perceived "red" is considered "blue for A" to B, there is no problem with A and B's claim that "the setting sun is red." This "inverted qualia" thought experiment argues that the idea of physicalism cannot explain differences in internal experiences, such as the appearance and perception of colors among observers.

### 1.4 Existence of free will

B. Rivett attached a device that measured brain electrical signals to participants in the experiment and had them move their wrists freely to measure brain electrical signals at that time (B. Rivet, 1980). When we move our bodies, it is known that electrical signals, commands to move our muscles, come out of our brains shortly before we move. The results of the experiment were in the following order.
1)The brain produces electrical signals to move the wrist
2)Conscious to move the wrist
3)Wrist moves

This means that the brain is sending electrical signals to move the wrist before the person is conscious of moving the wrist, and the wrist is already set to move before the person is conscious of moving the wrist. In other words, the wrists did not move because he was conscious of moving them. The normal human feeling that the wrist has moved because he or she wanted to move the wrist by his or her own will is an "illusion," and before that will, the wrist has been determined to move by something unconscious, and the command is coming out of the brain unconsciously. In the middle of the process of moving the wrist, it is a state in which the conscious mind is aware in the pursuit of "let's move." B. Rivet, between 2) and 3) above, there is a short time left for refusing to move the wrist, within which man can stop the action at his own will. So, he concludes that human beings have free will. However, there is no harm in interpreting the results of this experiment by thinking that it is "consciousness" - i.e., that there is no free will - to recognize the results of unconscious random processing in the brain as an afterthought. B. Rivet's interpretation of the experiment has been criticized by dualistic interpreters and others. However, if we take the view of "consciousness" in the absence of free will, we can assume that the brain unconsciously and randomly processes certain options based on physicalism. If we consider the post-processing recognition at the stage when the processing is determined to be "consciousness," we believe that it does not go against physicalism and does not cause a "hard problem of consciousness.". In dualism, apart from physical phenomena, there is non-physical "consciousness" that cannot be measured by physics. However, it fails to explain why and how non-physical consciousness can influence our behavior and the processing of the brain, which is a physical phenomenon. If consciousness, which is non-physical, can affect physics, we must also consider the existence of so-called "telekinesis" and the possibility of consciousness in all things, which seems to make interpretation more complicated.

### 1.5 Isolated Brains

Isolated brain is a condition in which the corpus callosum connecting the left and right sides of the brain is removed for the purpose of treating epilepsy, etc. This prevents information from being exchanged between the left and right sides of the brain. The left side of the brain is primarily responsible for language areas, while the right side is responsible for imaging areas. If we ask a subject with a separate brain, showing the subject only to his right eye, which is processed by his left brain, "What is it?" he can tell. Conversely, if the left eye, which connects to the right side of the brain, is seen alone, the right side of the brain, which connects to the right side of the brain, answers "I can't see anything" because it can't process language. On the other hand, if we let him draw a picture of the subject, his right brain, which controls the image area, can draw without any problem. It is sometimes said that this separate brain subject has two personalities = consciousness. Considering this, we think that consciousness is physically dominated, or at least influenced, by the physical state of the brain.

## 1.6 Short-term and Long-term Memory

Memories are divided into short-term and long-term memories because of differences in their temporal duration. Tulving further subdivides this long-term memory into "episodic" and "semantic" memories. "Episodic memory" is a memory of a personal experience associated with a specific time and place, and "semantic memory" is defined as a memory unrelated to a specific time and place (Tulving, 1976). As a classification of long-term memory, "episodic memory" and "semantic memory" may be combined into "declarative memory" and classified into two separately defined categories of "procedural memory" (Squire, 1987). Procedural memory, for example, is said to be the memory of how to ride a bicycle.

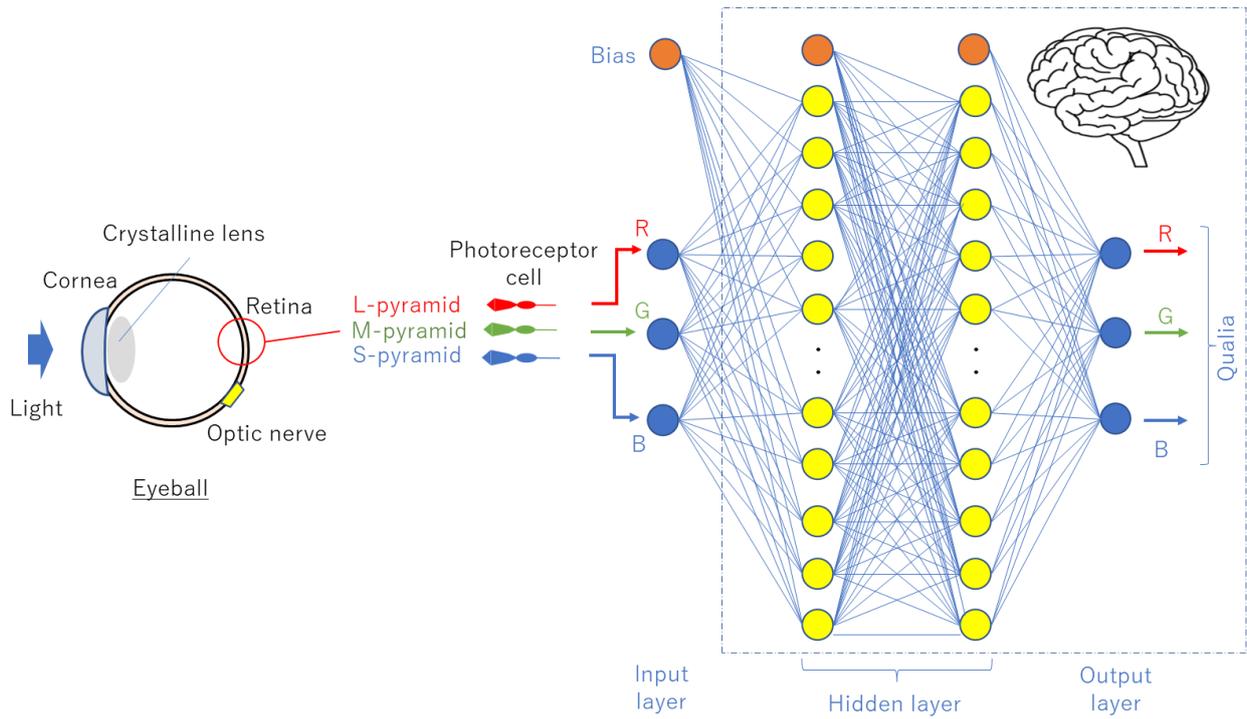

FIG. 1: Neural Network Configuration

Table 1: Color representation by RBG

| No. | Reference colors | | R | G | B |
|---|---|---|---|---|---|
| 1 | Red | | 255 | 0 | 0 |
| 2 | Green | | 0 | 255 | 0 |
| 3 | Blue | | 0 | 0 | 255 |
| 4 | White | | 255 | 255 | 255 |
| 5 | Black | | 0 | 0 | 0 |

Table 2: Learning results when Table 1 is used as input and output teacher signals

| No. | Simulated Colors | Input | Output | R | G | B |
|---|---|---|---|---|---|---|
| 1 | Red | | | 229 | 33 | 36 |
| 2 | Green | | | 38 | 227 | 38 |
| 3 | Blue | | | 38 | 39 | 231 |
| 4 | White | | | 254 | 255 | 253 |
| 5 | Black | | | 0 | 0 | 0 |

Table 3: Learning results considering color weakness

| No. | Simulated Colors | Input | Output | R | G | B |
|---|---|---|---|---|---|---|
| 1 | Red | | | 124 | 128 | 17 |
| 2 | Green | | | 124 | 115 | 18 |
| 3 | Blue | | | 185 | 202 | 255 |
| 4 | White | | | 255 | 255 | 255 |
| 5 | Black | | | 2 | 2 | 1 |

Table 4: Learning results when complementary colors in Table 1 are used as teacher signals for output

| No. | Simulated Colors | Input | Output | R | G | B |
|---|---|---|---|---|---|---|
| 1 | Red | | | 29 | 238 | 70 |
| 2 | Green | | | 245 | 17 | 62 |
| 3 | Blue | | | 236 | 161 | 32 |
| 4 | White | | | 255 | 255 | 255 |
| 5 | Black | | | 2 | 1 | 0 |

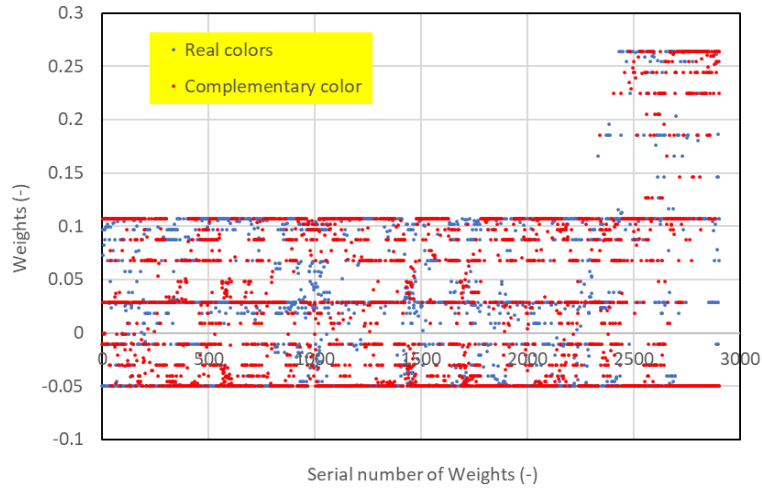

FIG 2: Comparison of neural network weight coefficients

Many existing models of consciousness seem to focus on first-person consciousness. In this paper, we first think of the neutral network as a Toy Model of a brain, and we performed a simulation of the inverted qualia using it. As a result, we show that color blindness and inverted qualia can occur even in the same network structure only with different values and arrangements of coefficients. On that basis, we defined second-person consciousness - especially qualia - mathematically and used the results as axioms to examine the process of creating a physically possible consciousness. We aim to build a model that can provide some answers to the above 1.1－1.6 questions. Here are the details:

## 2.Mathematical definitions and modeling of consciousness
### 2.1 Simulations of Inverted qualia

The possibility of 'inverted qualia' mentioned in the introduction is explored through a simulation based on a neural network. Neural networks, a type of machine learning that mimics the function of neurons in the brain, have many applications and achievements as so-called artificial intelligence. Now consider the three primary colors of red (R), green (G) and blue (B) and the five colors of white and black. Using RGB, each color can be quantified as a combination of RGB as shown in Table 1. The colors in the table consist of the RGB color in the right column. This simulation explores qualia as a first-person. First, the RGB values for each color in Table 1 are taken as input, and the same RGB values are used as the teacher signals for each color's output. In the model of the eyeball shown in the left part of Figure 1, the optic nerve contains a red-sensing L-cone, a green-sensing M-cone, and a blue-sensing S-cone, each of which transmits RGB signals to the brain. The input in the neural network is interpreted as the RGB value from the optic nerve of the color seen by the eye, and the neural network is interpreted as the processing in the brain and the RGB value of the output as the color-qualia recognized in the brain. The calculations were performed with input layer 3 nodes, hidden layer: 2 layers × 50 nodes, output layer 3 nodes, and a neural network considering bias at each node.

The weighting factors defined on the edges connecting each node were optimally calculated with MOST, which we developed (S. Inage, 2022). The search area for the weighting factor is -0.05−10. The total number of nodes is 2903. The study results are summarized in Table 2. The colors of Input in the table are those using the RGB values in Table 1, and the colors of Output in the table are those created from the RGB values listed in the right column. Overall, it can be reproduced with input color = output color. However, the RGB values do not exactly match those in Table 1.

Next, for the purpose of simulating so-called color weakness, when red RGB in the color weakness condition is recognized as = (128, 140, 0) (■)) and green is given RGB = (128,128, 0) (■) as the output teacher signal, the results are shown in Table 3. Other colors - blue, white, black - give identical RGB. The inputs remain in Table 1. Again, the colors of Input in the table are those using the RGB values in Table 1, and the colors of Output in the table are those created from the RGB values listed in the right column. From Table 3, red and green are generally more greenish compared to Tables 1 and 2, making it difficult to distinguish between red and green - that is to say, a state of color weakness. The color tone is also close to that shown in the reference above. Blue is also a lighter shade compared to Tables 1 and 2, but this is likely to be solved by learning more deeply about neural networks (blue is recognized as blue even when the color is weak). In this way, we can see that by changing the weights of the neural network, we can reproduce the normal state to the color weak state.

Next, consider a case where the RGB values of the inputs remain in Table 1 and the complementary colors of the colors in Table 1 are given as the teacher signals for the outputs. The complementary color of red is green (R, G, B) = (0, 255, 0), the complementary color of green is red (R, G, B) = (255, 0, 0), and the complementary color of blue is orange, (R, G, B) = (233, 163, 0). White and black are not colors and cannot define complementary colors, so they are left as they are. For machine learning, when a certain color is input, it can be regarded as a process to find the complementary color. The results are shown in Table 4. As above, the colors of Input in the table are those using the RGB values in Table 1, and the colors of Output in the table are those created from the RGB values listed in the right column. Again, the complementary color relationship to the input is correctly determined (except for black and white). In the calculations in Tables 2 and 3, the neural networks used are identical, only with different weighting factors on the edges connecting each node. Each weighting factor is a mathematically and physically feasible combination. This means that even with the same input, the same system can achieve different color outputs - equivalent to qualia - under different weighting conditions. Even if observer A, which has a weight coefficient outputting Table 2, and observer B (inverted qualia), which has a weight coefficient outputting Table 4, observe the same color, there is no contradiction even if they feel different qualia. That is, red for A is green for B, and green for A is red for B. But if we ask both parties what color they see, they can answer "red" or "green" and share it with each other. This is truly an inverted qualia. In the network of neurons in the brain, in the processing in the individual's brain, in the network, it is unlikely that there is a mechanism for perfect conformity. For example, each value of the neural network weighting factors of 2903 in each case in Tables 2 and 4 is compared in Figure 2. Up to a factor of 1-2300, the coefficient values are distributed in the range of -0.05－0.12 in all cases, and the coefficient values up to the latter 2300－2903 are distributed in the range of -0.05－0.27, and it is not observed that the coefficient values are significantly different, only that the distribution of the weighting factors changes. Because neutral networks are the Toy Model of the brain, but can also occur in that Toy Model, it is more natural to assume that in a more complex brain, individual qualia when observing things are different - even inverted qualia is possible. In this case, a qualia that is identical for everyone cannot be defined uniformly and is difficult to deal with mathematically. Giving up the discussion of qualia as a first-person, and avoiding its difficulties, we consider a framework that can be handled mathematically in the next section.

## 2.2 Definition of public language

Individual "qualia" is called qualia as a private language, after Wittgenstein's "private language" that conveys individual feelings and sensations. If this qualia as a private language are individual, different from each other, and incapable of communicating 100% to others, it is difficult to set a standard for mathematical comparison. Therefore, in contrast to "private language," the standard "public language" is defined from the qualia as a private language. In other words, it is a language in the second and third person. Take that definition and create a "standard" of consciousness.

First, assume that different observers A and B observe the same event F. Observer A recognizes the probability distribution of an event F as $P_A$, and observer B recognizes the probability distribution of the same event F as $P_B$.

In the probability space $(\Omega, F, P)$ and the probability variable X, consider the probability space $(\Omega, F, P_A)$ and the probability variable X for observer A, and the same probability variable X as the probability space $(\Omega, F, P_B)$ for observer B. In the probability space, P satisfies the follows:
  a) $P(\Omega)=1$
  b) $P(\emptyset)=0$
  c) $A \cap B = \emptyset$ for A, B $\in$ F, then $P(A \cup B)=P(A)+P(B)$

Also, F is σ additive family, which satisfies the following conditions:
  a) $\emptyset \subset F$
  b) $A \in F$ then $A^C \in F$
  c) $A_i \in F$、$i \in N$ then $U_{i=1}^{\infty} A_i \in F$

For the above two probability distributions $P_A$ and $P_B$, let us assume that we have density functions $p_A(x)$ and $p_B(x)$ defined below.

$$P_A(L) = \int 1_L(x) \cdot p_A(x) \cdot dx, \quad P_B(AL) = \int 1_L(x) \cdot p_B(x) \cdot dx \qquad 1)$$

where,

$$1_L(x) = \begin{cases} 1 \ (if \ x \in L) \\ 0 \ (if \ x \notin L) \end{cases} \qquad 2)$$

and is the defining function of the set L. In this case, the Kullback–Leibler divergence $D_{KL}$ is defined below.

$$D_{KL} = \int p_A(x) \cdot log\left(\frac{p_A(x)}{p_B(x)}\right) dx \qquad 3)$$

This is one expression of the distance between $P_A$ and $P_B$ and satisfies the following conditions:
  a) $D_{P_A P_B} \geq 0$ for any $P_A(x), P_B(x)$
  b) $D_{KL} = 0$ and $(\forall x) P_A(x) = P_B(x)$ are equivalent
  c) Generally, $D_{KL}(P_A, P_B) \neq D_{KL}(P_B, P_A)$

If the distance is zero, the following holds:
  a) Reflectance A~A
  b) Symmetric Toru A~B
  c) A ~C if transitive laws A~B, B ~ C
where, "~" is a symbol denoting equivalence.

We define this as the 'equivalence' of the results for an 'event' between observer A and observer B. In this paper, "events" indicating this equivalence are collectively referred to as "public language." I think this idea is close to the relationship between private and public language in Wittgenstein's "Private Language Theory." Private language refers to language that expresses sensations, emotions, and other things that can only be understood by the person. Wittgenstein believed that "private language," which conveyed individual feelings and sensations, was meaningless, becoming "public language" when conveyed to those around them. Therefore, in the theory of private language, Wittgenstein concludes that 'private language is inherently unmastery and meaningless. The above probability distribution by individual observer corresponds to "private language," and if the distance of the probability distribution is zero, it becomes "public language." This is illustrated by the case of colors calculated in the previous chapter. For simplicity, the three colors are {Red, Green, Blue}. In this case, the sample space $\Omega$ is {Red, Green, Blue} and F: $2^{\Omega}$={∅, {R}, {G}, {B}, {R, G},{R, B}, {G, B}, {R, G, B}} is a power set of $\Omega$.

It is assumed that observer A sees the color with the RGB reference values in Table 2. The $F_A$ in that case is as follows:
  $F_A$ ={∅, {229, 33, 36}, {38, 227 38}, {{38, 39, 231}, {{229, 33, 36}, {{38, 227 38}}, ・・・}

Suppose that observer B sees color according to the criteria in Table 4. The $F_B$ in that case is as follows:
  $F_B$ ={∅, {29, 238, 70}, {245, 17 62}, {{236, 161, 32}, {{29, 238, 70}, {245, 17 62}, ・・・}

Thus, even if the criteria for each color are different, if we observe colors based on those criteria, the probability of colors will match. In this case, we calculate the probability density function, the Kullback–Leibler divergence is zero, and the observed colors can be considered equivalent in A and B.

Also, σ additive family: F contains an empty set. In the case of color, it can be regarded as not feeling color - that is, not feeling color qualia. Considering this as a philosophical zombie state, the framework of this proposal allows and encompasses philosophical zombies as elements. On the other hand, in the probability space (Ω, F, P), since P (∅) = 0, the philosophical zombie can be an element, but it can be interpreted that the probability of the philosophical zombie is zero in the public language. This framework considers that defining public language avoids the philosophical zombie criticism of physicalism.

In this paper, the causes of the differences that arise in qualia as a private language cannot be identified, as in ordinary probability theory. The following two possibilities can be considered as qualia.
1) There is a "consciousness" that produces identical qualia—this is a dualistic position.
2) There are no qualia as a "universal" private language, only a public language derived from an individual private language qualia. Individual qualia — as qualia as a private language, encompassing inverted qualia, even those without qualia, such as philosophical zombies. Qualia as a private language, like individual events in probability theory, cannot discuss differences individually. However, public languages can be handled mathematically.

In the position of 1) above, we immediately face a "hard problem of consciousness." However, in the position of 2), even if the mechanism by which differences in individual qualia occur cannot be grasped, it is possible to define public language and perform mathematical operations by treating it as a probability space and a random variable, as described above. Even though there are individual intrinsic qualia - qualia as private language - public language is essential to the initial learning process. In other words, for an infant to acquire a qualia of "red," it is essential that the public language be provided by both parents and relatives and that learning be based on the recognition of their equivalence. While the position that "public language" is mathematically definable and computable is physicalist, it does not touch on qualia as a private language. This is the case in the question of probability, even when a coin is tossed and a table appears, if the detailed process of the initial conditions, the amount of force in the toss, etc. is followed by Newtonian mechanics, why the table appears is perfectly possible. On the other hand, even if the details are skipped, it is like the situation in which a mathematical deduction is possible using the event of a coin flip as a probability. From the above discussion, I think it is possible to develop a physicalism that mathematically deduces, at least for public languages, their nature - consciousness as a public language. The merit of this introduction is, rather, obvious because it presupposes a comparison with a third party, but the public language is capable of third-party measurement. Cases with inverted qualia, philosophy zombies, all ascribe to the same public language. They can also tell green from red through public language - so-called color weakness. If this third party introduces measurable parameters and models consciousness, a mathematically deductible theory should be constructed.

In summary, we believe that the qualia as a private language can define a public language by observing the same events with each other and making them common - for example, in language, etc. In this paper, we will proceed with the discussion by taking it as an axiom that this public language can be defined. In this proposal, the generation of consciousness is considered to occur in the following STEP -1 to STEP -4. The following sections describe each STEP.

## 3. Modeling of will and consciousness

In this paper, creation process of consciousness is classified into four steps. Namely, STEP-1: Recognition by external perception, STEP-2: Connection with past episodic memory, STEP-3: Decision making, STEP-4: Recognition by internal perception. Each step is explained below.

### 3.1 STEP-1: Expressions of 'recognition' by external perception and synthesis between individual qualia - creation of episodes

In the Kullback–Leibler divergence defined from multiple functional spaces $(\Omega, F, P_i)$, if the distance is zero, it can be defined in terms of individual official languages. Next, we describe a more complex definition of official language. For example, combining individual official languages - combining "white" and "dog" to create "white dog." The qualia of this combination can be expressed naturally if one considers that the sample space, ohm, is the direct product of the sample space, $\Omega_C$, for color and the sample space, $\Omega_A$, for animals: $\Omega = \Omega_C \times \Omega_A$.

In that case, the element of $\Omega$ would be [White, dog], [Black, cat] ・・・, etc. It is self-evident that from this sample space, if we define, for example, a power set $F = 2^\Omega$, we can generate a $\sigma$ additive family and its probability space.

Its mathematical definition is as follows:

Consider two probability spaces: $(\Omega_j, F_j, P_j)$ (j = 1,2). We define these Cartesian product probability spaces $(\Omega, F, P)$ as follows:
  a) We first define $\Omega$ as the direct product set of $\Omega_1$ and $\Omega_2$.
     $\Omega \equiv \Omega_1 \times \Omega_2 \equiv \{ (\omega_1, \omega_2) \mid \omega_1 \in \Omega_1, \omega_2 \in \Omega_2 \}$
  b) F is defined in stages as follows
     1) $C \equiv \{A_1 \times A_2 \mid A_1 \in F_1, A_2 \in F_2\}$
     2) A is the disjoint original finite sum of C.
     3) F is the smallest $\sigma$-additive family containing A: i.e., $F = \sigma(A)$.
  c) Finally, the probability P is:

1) For the set of forms $A_1 \times A_2$ $(A_1 \in F_1, A_2 \in F_2)$, $P[A_1 \times A_2] \equiv P[A_1] \cdot P[A_2]$ is defined.
2) The more general element of F (For example, a set of direct products of the form $A_1 \times A_2$, or an 'extreme' set because F is a σ-additive family, etc.) is extended and defined by imposing σ-additivity on P. Namely, let $P(A \cup B) = P(A) + P(B)$, $A \cap B = \emptyset$.

Furthermore, for example, if actions are expressed in gerund terms and rules are defined such as "verb" →"adjective-1"→"adjective-2"→"object," ・・・ more complex qualia such as "running white and big dog" can be generated, and conversion such as "White and big dog is running" can be easily done. This processing creates an observation-based "episode" as a direct product of the current public language. Episodes don't necessarily mean writing, but they include extracting only simple images from actual observations, such as "dog" or "white," and compositing them into minimally simple images. So, to speak, it is supposed to remove noise and extract only the "essential parts necessary" and express them as internal sentences, images, sounds, tactile sensations, etc. This created episode also has the character of a public language. Among the above episodes, a sentence episode is a sentence episode, and an image episode is an image episode. If we think of them as being produced individually in the speech and image areas, this representation offers suggestions for understanding separate brains. It is said that the left side of the brain controls the language area, and the right side controls the image area. If we split the corpus callosum that connects them, we can't create a direct product of left-brain speech episodes and right-brain image episodes that are generated by information from the left and right sides. In other words, it is impossible to create an episode that integrates information from the left and right sides, and the left brain makes decisions by episodes of the language area, while the right brain makes decisions by episodes of the image area. The proposed expression can encompass the phenomenon that a person with a separated brain can draw a picture of an object seen with his left eye but cannot explain it in words, and conversely, he can explain an object seen with his right eye but cannot draw it.

The moment at which this external perceptual perception arises can be mathematically represented as the difference in the average information entropy before and after episode creation. Before recognition, the average information entropy is zero, since $-\sum P_i log(P_i)$ has a finite value, whereas after recognition, it aggregates into one episode. It is natural to define this moment of change from a finite value to zero as the moment of "recognition by external perception."

As described above, since episodes generated by combining individual official languages are defined as a probability space, we believe it is possible to observe and confirm them by a third party by introducing and comparing the Kullback–Leibler divergence $D_{KL}$.

Finally, the technology for creating sentence episodes - so-called tags - from information such as photos is already called annotation and has already been implemented in various applications (Papadopoulos, Dim P., et al, 2017).

### 3.2 STEP-2: Associate Current Episodes with Past Episodes

It is assumed that episodes created based on past observations go through a short memory and important ones are stored as episodic memories. Past episodic memory should also be represented as a sample space as a direct product between the individual public languages mentioned above. We think of this sample space as an accumulation of past experiences and a set with enormous elements.

Indeed, past episodic memories can be written out and compared by third parties.
In this section, we consider, in the past episode, the feelings and actions of experiences. Emotions are also not individual feelings as a private language, but only feelings as a public language, as mentioned above. In this paper, as a basis, we consider the sum of Plutchik's emotions (Plutchik, 1982). Plutchik proposed that all emotions, like the tri primary RGB of light, are formed using eight basic emotions (primary emotion, called pure emotion): joy, trust, fear, surprise, sadness, disgust, anger, and expectation. For example, "love" as a secondary emotion is a mixture of "joy" and "trust," and "curiosity" is a mixture of "trust" and "surprise", etc. If Ω = {Joy, Trust, Fear, Surprise, Sadness, Disgust, Anger, Expectations} and the state of not feeling any emotion is an empty set, then it is obvious that a probability space can be defined, as in the example of color, etc. It is also clear that this probability space contains elements such as the above examples: {joy, trust}: love and {trust, surprise}: curiosity. These emotional episodes are stored as episodic memories, in the form of sentences and images. If the current observations provide a textual episode of "Walking white and big dog," we can discuss the probability of each emotion occurring when we recognize "Walking white and big dog" in the probability space. For example, joy: 0.5, fear: 0.2, disgust: 0.1, etc.

If the probability is non-zero, it can be addressed as past feelings related to the current episode. In addition, emotion-related behaviors - such as running away from fear or approaching from joy - can also be retrieved from past episodic memories. The sample space then becomes a representation as a direct product of the sample space of emotions and the sample space of behavior. With this operation, multiple episodes of feelings, experiences, etc. experienced in the past can be associated with events from current observations. Consider that an episode with a non-zero probability is recognized as relevant to the present and a past episode with a zero probability is not recognized.
In summary, for an episode of "running white dog", it is a probability space consisting of a sample space of several probabilistic events: for example, in the past, emotional episodes such as "bitten," "barked", "enjoyed", and "cute", for "running white dog", etc. For a "running white dog", consider an operation in which multiple high-probability items are selected from elements in the probability space. "High probability" is the most "impressive" if it is captured with feelings. From this operation, by relating episodes from current observations to past episodic memories, we can select multiple events from past episodes that should occur in the future.

### 3.3 STEP-3: Modeling Decision-Making
One of the important roles of consciousness is thought to be related to survival. That is, how should we act on the present event, then - i.e., the future - based on experience? From now on, the "role" of consciousness will be to take out "past experiences" based on "present information" and determine the next "future action." We believe that one of the simplest measures is to follow past performance-successful experiences. In other words, as mentioned in the previous section, from past episodic memory and current information, the "future" chooses what action to take. In STEP-2, there are multiple past episodes related to the present observational event. The actual behavior of the future shall be chosen entirely at random or in Markov chains from multiple past episodic memories and the associated behavioral experiences - performance - i.e., by chance. This randomness is determined by the physical state of the brain. For example, differences in the electrical potential pulses of neurons, mistimed neurotransmitter release, etc. This is called "will," and based on it, actual action is taken. This idea makes it possible to account for current and past episodic information, as well as the creation of multiple alternatives from past episodic memory, and selection as a physical phenomenon. There is no "free will" involved in this selection process, and the unconscious = physical phenomena in the brain, including noise - selects at random. That is, this manipulation can be explained in terms of physicalism. We consider a mathematical model of " random will choice". To do this, we first consider "will" as an expectation. As mentioned above, the framework of this proposal is that the choice of action - the will - for the future is made randomly from the sample space of the past qualia. The choice is chosen from the probability space of the past episode. The probability space is accompanied by expectations. When we say "will", we mean our own choice. In other words, if the reality is to be done stochastically, the object must have said "will" to the expected value from the probability space. First, we focus on will as an expectation, and for modeling purposes, we define the

"hypothetical dynamics" of will. We think of volition as the process of changing the initial state probability space to create a new state probability space. First, we define position x as the Kullback–Leibler divergence $D_{KL}$ between the probability space representing the initial state and the future probability space realized by the will. This distance is not limited to the Kullback–Leibler divergence, but may be any other distance. We believe that $D_{KL}$ is a measure of the difference between probability distributions, and if we follow its time change, it can be perceived as a change in position in normal dynamics. In addition, the time derivative of x corresponds to velocity and can be interpreted as the speed of change in the probability distribution, and the second derivative of time can also be considered as acceleration α. Furthermore, consider, perhaps intuitively, the "inertia" of the will. There is a lag in remembering when we must remember something and act on it. We will call this degree of slow speed "inertia" and that degree the mass m of decision making. The speed and acceleration defined above are the processing speed to the decision after the recall, and are considered as independent parameters of m. In this paper, we consider decision making as a selection from past behavioral information from past episodic memory, and the acceleration α, mass m defined above should be closely connected to past episodic memory. Then, again in analogy to dynamics, we define the potential V of past episodic memory as a function to satisfy:

$$-\frac{\partial V}{\partial x} = m\alpha \qquad 5)$$

Now let us assume that m×α represents a force in ordinary dynamics, but within Equation 5), we understand it as a relation defining the potential V, and there is no further meaning. It is natural to think that past episodic memory influences the speed and acceleration of the will and introduces potential as an indicator. All the above definitions of hypothetical dynamics are for will as expectation. In contrast, we think that in actual decision making, randomness is added by the state of the brain - the state of neurons, etc. That is, in Brownian motion, the same image as fine particles being shaken by surrounding liquid molecules. In fact, the velocity of Brownian moving particles varies randomly, so that dx/dt cannot be defined for each instant. Similarly, in decisions involving randomness that is not an expectation, it is impossible to define such things as speed as the expectation mentioned above.

Therefore, we define the time variation of position x as follows.

$$x(t + \Delta t) - x(t) = b(x(t), t)\Delta t + w(t + \Delta t) - w(t) \qquad 6)$$

where w(t) is the Wiener process. This represents that the time variation of the Kullback–Leibler divergence changes under the influence of the Wiener process. Based on 6), consider calculating the velocity and acceleration in decision making as expected values. So, when there is a random variable f (t) that generally depends on time t, its < forward mean derivative > is defined below.

$$D[f(t)] \equiv \lim_{\Delta t \to 0} \left\langle \frac{f(t+\Delta t)-f(t)}{\Delta t} \middle| f(s) \ (s \leq t) \right\rangle \qquad 7)$$

The < | > on the right-hand side represents the conditional time average that f (s) before t is fixed. In this 6), the forward average differential coefficient of x (t) is obtained as follows:

$$D[f(t)] \equiv b(x(t), t) \qquad 8)$$

Next, < Backward mean derivative > is defined below.

$$D_*[f(t)] \equiv \lim_{\Delta t \to 0} \left\langle \frac{f(t)-f(t-\Delta t)}{\Delta t} \middle| f(s) \ (s \leq t) \right\rangle \qquad 9)$$

Using Eq. 9) to determine the backward-facing mean derivative of x (t), we obtain:

$$D_*[f(t)] \equiv b_*(x(t), t) \qquad 10)$$

Using this b*, we obtain:

$$x(t) - x(t - \Delta t) \equiv b_*(x(t), t) + w(t) - w(t - \Delta t) \qquad 11)$$

The acceleration α at the expected value is then defined below.

$$\alpha(t) = \frac{1}{2}(D_*D + DD_*)x(t) \qquad 12)$$

Calculating the right-hand second term DD* of the above equation, we obtain:

$$DD_*x(t) = \frac{\partial b_*}{\partial t} + b\frac{\partial b_*}{\partial x} + \frac{v}{2}\frac{\partial^2 b_*}{\partial x^2} \qquad 13)$$

Similarly, D*D is:

$$D_*Dx(t) = \frac{\partial b}{\partial t} + b_*\frac{\partial b_*}{\partial x} + \frac{v}{2}\frac{\partial^2 b}{\partial x^2} \qquad 14)$$

Next, the following variables are introduced:

$$u = \frac{1}{2}(b - b^*), v = \frac{1}{2}(b + b^*) \qquad 15)$$

Using this, the expected acceleration $\alpha$ is:

$$\alpha = \frac{v}{2}\frac{\partial^2 u}{\partial x^2} - v\frac{\partial u}{\partial x} + u\frac{\partial v}{\partial x} + \frac{\partial v}{\partial t} \qquad 16)$$

Applying 5) to this $\alpha$, we obtain the following relation:

$$\frac{\partial v}{\partial t} = \frac{v}{2}\frac{\partial^2 u}{\partial x^2} - v\frac{\partial u}{\partial x} + u\frac{\partial v}{\partial x} - \frac{1}{m}\frac{\partial V}{\partial x} \qquad 17)$$

If the above definitions are allowed, the theory that E. Nelson tried to explain the governing equations of quantum mechanics based on the Wiener process is applicable as is. A summary of the derivation process is provided in Appendix. Only the conclusions of the governing equations of the probabilistic process of will obtained since the above assumptions are shown below.

$$iv\frac{\partial \psi}{\partial t} = \left[-\frac{v^2}{2}\frac{\partial^2 \psi}{\partial x^2} + \frac{1}{m}V\right]\psi(x,t) \qquad 18)$$

The $\phi$ corresponds to the wave function of quantum mechanics, which shows that the 'decision' is stochastic and is connected with the distribution function $\rho$ (x, t) in the case of settling into one state by:

$$\rho(x,t) = [\psi(x,t)]^2 \qquad 19)$$

John Duffy and Ted Loch-Temzelides carried out experiments on decision making that correspond to the double slit in quantum mechanics and observed that decision making also has particle and wave duality (John Duffy and Ted Loch-Temzelides, 2021). The results of this experiment should be further examined and discussed, but since the model of decision making in this proposal is identical to the Schrodinger equation, it is concluded that the experiment of John Duffy and Ted Loch-Temzelides can be directed by Eq. 18). And the values and functions of the parameter m and potential V are unknown, but might be determined by comparison with the experiments of John Duffy and Ted Loch-Temzelides.

### 3.4 STEP-4: Recognizing with Internal Perception - Modeling Consciousness

In STEP-3 of the previous section, it is assumed that the actual action is done "unconsciously" and that the person observes an episode (a sentence or an image) created from the action result, which can be called "introspection". That observation, as described above, creates a current episode associated with one's own behavior (called recognition) and is remembered as a short-term memory. Think of it as a moment when recalling becomes possible - when information can be extracted from short-term memory. Let this moment be "the moment when consciousness comes to recognize that one has acted". This suggests that the "time of onset" of consciousness lags actual behavior. If this process of "will" and "generation of consciousness" were to take place in an extremely short time, one would not be aware of the two processes and would recognize that the actual "consciousness" had decided. This interpretation resolves the following:
1) The brain must be governed only by physical phenomena, and we do not know why "consciousness," which is not a physical phenomenon, can alter physical phenomena.

2) The existence of "free will" by B. Rivet, which seems to deny free will (B. Rivet, 1983).

As mentioned above, in this proposal, the "will" associated with the "present episode" and the related "past episodic memory" is randomly selected, which is consistent with what is done in the physical phenomena of the brain - i.e., consciousness does not change the brain's choice. What to choose from among the choices is purely accidental. If we consider the moment of "doing" the action according to the will and "recognizing the action" as a summary of it as the generation of consciousness, it does not contradict the above tasks 1) and 2). In particular, the smaller the time difference between action and recognition, the more likely it is that people will recognize that they made their own decisions. Experiments with B. Rivet estimate this time difference to be around 0.3 seconds. In addition, if they act according to their memories of past episodes, they should generally evaluate their perceived "will" as a correct judgment. This current information - obtaining information as a combination of public languages→extracting the "will for the future" from the collation with past episodes → selecting actual actions at random from the "will" from multiple past episodes → identifying the "recognition" and "memory" by confirming the results of the actions as "consciousness". This process is supposed to be continuous with respect to time, so that continuity of consciousness appears. The moment of birth of this internal perceptual recognition can also be expressed mathematically as the difference between the average information entropy before and after confirmation. Before confirmation, the average information entropy is zero, because it aggregates into a single episode, whereas before confirmation, $-\sum P_i log(P_i)$ has a finite value. It is natural to define this moment of change from a finite value to zero as the moment of "recognition by internal perception".

Consider the role of consciousness in this case. Consciousness makes no contribution to current behavior, but new episodic memories created from current behavior may work effectively for future behavior. In other words, more episodic memory = more options for the future, such as making up for the shortfall in existing episodic memory or strengthening specific memories from existing episodic memory. This allows us to make better choices. In that sense, consciousness does not have free will to act in the present, but it does have more freedom to change the future in the sense that it "gives us more options" regarding actions in the future. This is not against physicalism, as consciousness does not influence physics, but rather determines future behavior through the information of episodic memory. That is, the idea in this proposal is that memory mediates between consciousness and physics, which are non-physical.

## 3.5 Explained by Toy Model

STEP-1 to STEP-4 above is explained with a simple Toy Model. For example, consider two colors, white and black, and dogs and cats as animals. If the sample space for color is . $\Omega_{Color}$ = [White, Black] and the sample space for animal is $\Omega_{Animal}$ = [Dog, cat], the sample space of the direct product is $\Omega_{Color \times Animal}$ = [{White, dog}, {White, cat}, {Black, Dog}, {Black, Cat}]. Since there are 4 elements, the set F=[∅, [{White, dog}], [{White, Cat},..., $\Omega_{Color \times Animal}$], which should have 4 elements, has $2^4$ = 16 elements. Consider the probability space($\Omega_{Color \times Animal}$, F, P) that will be created. Each observer has a 1/4 chance of identifying the correct animal. If we observe an actual white dog and recognize it as a "white dog" (creating an episode as a sentence or image), the probability of being a "white dog" at that moment is 1. We consider this in terms of mean information entropy. The mean information entropy H(X) is defined below.

$$H(X) = -\sum_{i=1}^{M} p_i log_2(p_i) \qquad 20)$$

where $X$ is the random variable, $p_i$ is the probability that event i will materialize, and M is the number of events. We calculate the change in the average information entropy before and after the observation.

$$\text{Before observation}: H(X) = -\frac{1}{4} log_2\left(\frac{1}{4}\right) \times 4 = 2$$

$$\text{After observation} : H(X) = -1 \times log_2(1) = 0$$

From this, the observation changes the average information entropy from 2 to 0. The above is STEP-1. The moment when the average information entropy changes is regarded as the moment of "recognition" and the expression of "consciousness" by outside observation. Next, when observed in STEP-1 and recognized as a "white dog," the episodic memory associated with the white dog is recalled. Here, as a sample space, we consider the eight basic emotions of Plutchik introduced above. Namely, $\Omega_{Emotion}$=[ Joy, Trust, Fear, Surprise, Sadness, Disgust, Anger, Expectations]. The probability of each emotion produced by the recognition of a "white dog" is defined as $P_{Joy}$ to $P_{Expectations}$. This is STEP-2. Which of these will be realized (selected) depends on the physical state of the brain, noise, randomness, and according to Eq. 18). Hence, "free will" is not reflected in the choice. This becomes STEP-3. Finally, this choice is evaluated as an episode and short-term memory, or more importantly, episodic memory. Again, as in STEP-1 above, there is a change in the average information entropy before and after selection.

$$\text{Before obserbation} : H(X) = -[P_{Joy}log_2(P_{Joy}) + P_{Trust}log_2(P_{Trust}) + \cdots + P_{Exp}log_2(P_{Exp})]$$

$$\text{After obserbation} : H(X) = -1 \times log_2(1) = 0$$

This becomes STEP-4. In this case, too, from the point of view of information entropy, the moment of the birth of consciousness can be regarded as the moment when the average information entropy goes from a finite value to zero. This is because the moment when the mean information entropy goes from finite to zero is a distinct discontinuity point. In this case, we considered an episodic memory of the emotion that occurs when we see a white dog, but in fact we choose future behavior in conjunction with, for example, the behavior associated with "fear", such as escaping from the dog. This completes the sequence of 1. recognition and episodeization of the object of observation, 2. recall of the relevant past episodic memory, 3. random selection from the past episodic memory, 4. recognition by episodeization of the selection. In this proposal, this whole process is considered as "creation of consciousness" and the moment of recognition by two episodic episodes of observation and selection is considered as "birth moment of consciousness." The above is summarized in Fig. 4. If a third party looked at this process - indeed, the reactions and actions of the person who saw the "white dog" - the subject would clearly perceive that he acted with a will and could ask why. The subject will respond that he or she has acted on his or her own volition, such as escape, based on past experiences. I think it is fair to say that this is consciousness arising from the assembly-episodic memory of a public language. This proposal assumes a common official language by comparison with the third person, so it is a model of consciousness that assumes the second or third person, not the first person. Therefore, the position is that consciousness exists in oneself because one feels conscious in others.

Next, the way of thinking about the mind-body problem in this proposal is considered. There are three main ways of thinking about mind-body problems:

1) Interactionism
   : The idea that there are two very different kinds of things in the world, the mental and the material (dualism), and that they interact. The idea is that substances in the brain can be influenced by the world of consciousness and behave differently from the laws of physics (Fig. 4a).
2) Epiphenomenalism
   : This theory is physicalism, with the position that consciousness and qualia are only phenomena attached to the physical state of matter and have no causal effect on matter (Fig. 4b).
3) Parallelism
   : This theory holds that the world is one in which consciousness and physics are two very different things, and the two proceed in parallel without interaction. This is the so-called position of dualism (Fig. 4c).

The proposal is a mixture of 1) and 2) (Fig. 4d). Consciousness does not directly affect the physics in the brain. However, consciousness can influence "future" brain choices through memory. As mentioned above, consciousness in this proposal is an operation in which the 'action physically determined by the brain' is made into an episode and left as an episodic memory for the future. The present will be determined before consciousness occurs, so it does not reflect the present. However, the increased choice of episodic memory indirectly affects future activity. In other words, stored information links consciousness (which appears to be non-physical) with physics. If we apply the word "free will," it means more options for the future, not for the present. The proposed idea can also be seen as an evolutionary system of "reflexes" in which we act directly in response to stimuli without involving the brain. Whereas the stimuli in the reflex are limited, the brain accumulates the episodes it experiences, thereby allowing it to respond to more flexible stimuli. In other words, while reflexes limit stimuli, "consciousness" in this proposal means that stimuli can be rewritten. It seems natural to the author to think that consciousness was acquired in this process of expanding from limitation to generalization.

## 4. Discussions and Conclusions

The features of the proposed modeling of consciousness generation are as follows.
1) Qualia as a private language is unrecognizable. On the other hand, by introducing a probability space and a random variable for an event consisting of qualia, a public language can be defined by the Kullback–Leibler divergence to the probability space of multiple observers.
2) The family F of events in the probability space contains an empty set, which can be regarded as a philosophical zombie. But in official language, the probability is zero. In other words, it can be said that philosophical zombies are possible, but the probability of them being zero is the public language.
3) Episodes that combine multiple official languages are the direct product of the probability space of each qualia, and similarly can be discussed in the probability space.
4) Episodic memories created in the past that remain are also probability spaces.
5) Create the most likely future (episodes) from observation-based episodes and episodic memories. In addition, multiple choices are made from "episodic memory" for coping with the future.
6) One of the above options for the future is chosen at random and executed "unconsciously." Randomness simply depends on the state of physical phenomena in the brain. The process up to this point is purely physicalism.
7) 6)conduct 1) - 3) above as an observation result, and the point at which an episode of an action is generated and stored in short-term memory - a point at which it can be recalled at any time - is defined as the "moment when consciousness is born." Important memories are moved from short-term memory to episodic memory.
8) 1) to 7) are time continuous, and in that sense consciousness is also time continuous.

We believe that these 1) to 8) flows can create models of consciousness based on mathematically definable public languages. We consider a comparison between our proposal and integrated information theory by Tononi. Integrated information theory is a position where qualia exists as a private language. We turn a blind eye to the hard problem of consciousness and, given the existence of consciousness, we construct a theory with axioms concerning its' information ',' integration ',' structure 'and' exclusion '. Then, the integrated information theorem, $\Phi$ is introduced, and the level

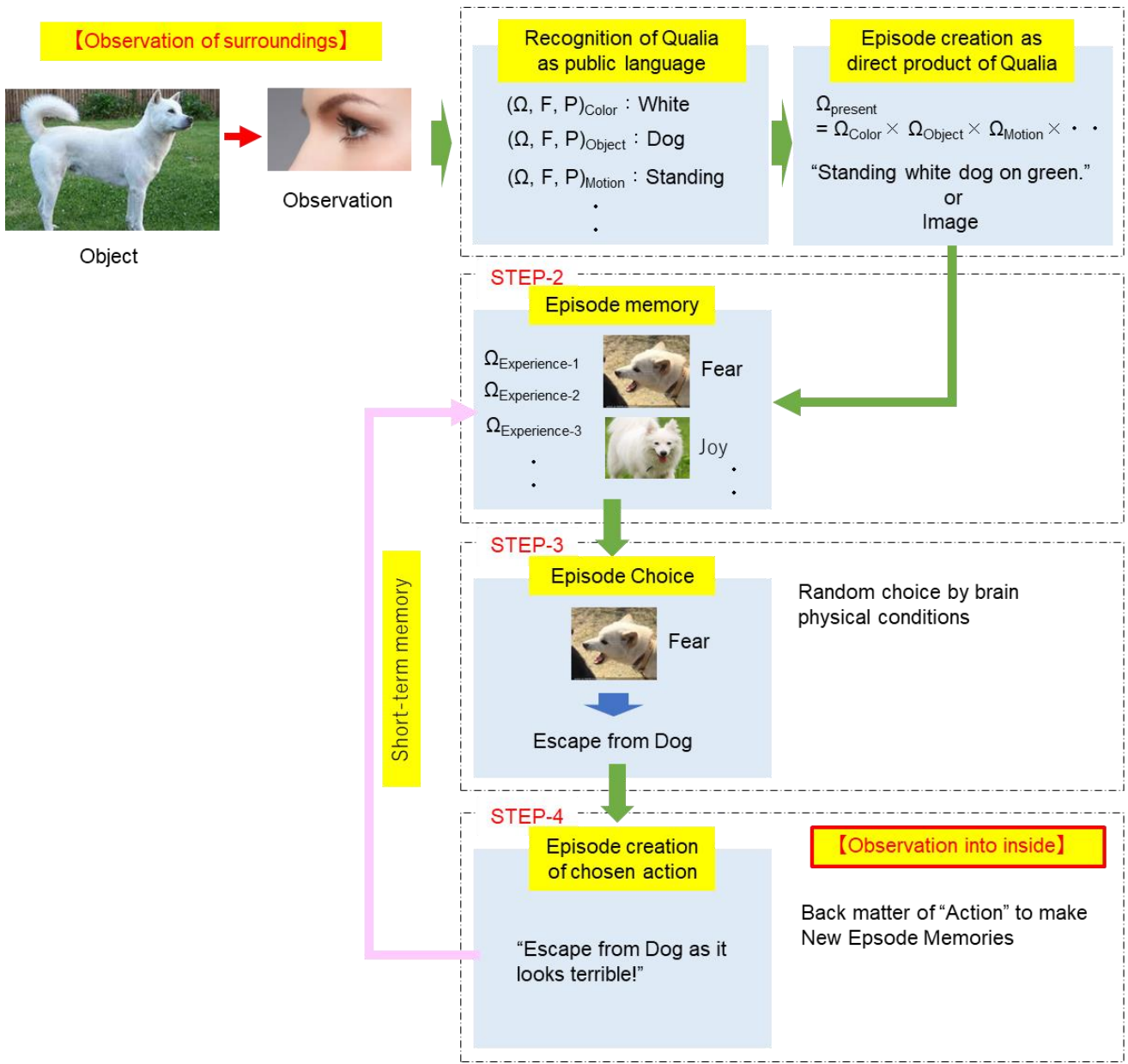

FIG. 3: Flow of creating consciousness in this proposal

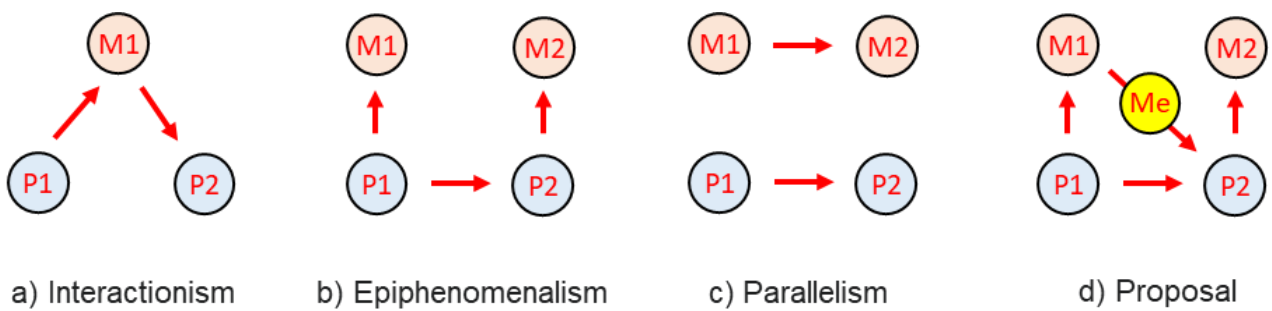

(M: Mind, P: Physics, Me: Memory)
FIG. 4: Comparison of ways of perceiving mind-body problems

of consciousness is quantitatively evaluated by the magnitude of the value of $\Phi$. According to integrated information theory, a digital camera, for example, has a huge amount of information depending on the pixels, but because the information is not integrated, the position is that we cannot have the visual awareness that we wait for. As mentioned above, since integrated information theory targets individual consciousness, it cannot be directly compared with ideas that assume public language such as this proposal. According to this proposal base, if we interpret the example of a digital camera, the digital camera cannot construct current episodes from the shooting information, nor does it have past episodes. So, it doesn't make comparisons between current episodes and past episodes, or even future behavioral episodes. Furthermore, they don't have the "inside view" of their chosen future. Therefore, digital cameras must be unconscious.

Next, as another philosophical topic, we will consider, through this proposal, the "Mary's Room," which is a counterargument to the physicalism advocated by F. Jackson (Jackson, Frank, 1982). The contents of "Mary's Room" are as follows. Mary, a brilliant scholar, has spent her life in black-and-white rooms since birth, gaining knowledge only in the black-and-white medium, and by nature has never been exposed to actual color. But they understand all the information about color and vision - the structure of the eye, the color is made up of RGB combinations, every object has a color, they fully understand that color as an RGB value, etc. Does this Mary learn anything new about vision when she goes out for the first time and is exposed to real colors? This is the question in this issue. The position of this proposal is that the qualia as a private language, and the public language should be clearly classified, and that the qualia as a private language cannot be measured while the public language can be measured. According to this, Mary would have learned exactly the "public language" mentioned in this proposal, so she would have learned nothing by touching actual colors. Of course, we believe that she may "feel" qualia as a private language when she touches color for the first time, but this proposal does not cover qualia as a private language, so she cannot get no learning in public language more than her own learning.

Next, brain waves are observed in brain cells cultured in the test tube - does consciousness reside in the so-called test tube brain? The premise of this proposal is that a public language formed by consensus among multiple observers is essential. In contrast, the test-tube brain has no means of observation and, by "appearance," cannot compare itself to others. Therefore, it is obvious that they cannot create and share official languages, and in that sense, they cannot have episodic memories that can be 'introspected', and therefore they will not have consciousness.

When a public language is defined in probability space, episodes, episodic memories, etc. that combine it can be defined without contradiction, and within that framework, consciousness can be understood as physicalism. They seem to dodge criticisms of physicalism: philosophical zombies, segregated brains, Mary's Room. As such, we believe that this proposal is both a self-fulfilling framework and a consistent explanation for the effects of consciousness we experience daily. Considering the above, this paper is not " Cogito ergo sum-I think, therefore I am" but "To feel a common self in the other person, therefore I am" because it considers consciousness as a public language. Also, the framework of this proposal may be sufficiently realized as software.

References

1) Plutchik, R. (1980). A general psychoevolutionary theory of emotion. In R. Plutchik & H. Kellerman (Eds.), Emotion: Theory, research and experience, Theories of emotion (Vol. 1, pp. 3–33). New York: Academic Press.
2) Plutchik R. (1982) A psychoevolutionary theory of emotions. Social Science Information. 21: 529-553. https://doi.org/10.1177/053901882021004003
3) "Basic Emotions--Plutchik". Personalityresearch.org. Retrieved 1 September 2017.

4) Chalmers, D. (1996): The Conscious Mind, Oxford University Press, New York.
5) Chalmers, David (21 March 2019). "Zombies and the Conceivability Argument". Phil Papers.
6) Kenny, Anthony (1973), Wittgenstein, Penguin Books, ISBN 0-14-021581-6.
7) Libet, Benjamin; Gleason, Curtis A.; Wright, Elwood W.; Pearl, Dennis K. (1983). "Time of Conscious Intention to Act in Relation to Onset of Cerebral Activity (Readiness-Potential) - The Unconscious Initiation of a Freely Voluntary Act". Brain 106: 623–642.
8) Velmans, Max (2000). Understanding Consciousness. London: Routledge. pp. 35–37. ISBN 0-415-22492-6
9) Dennett, D. The Self as Responding and Responsible Artefact Archived July 1, 2016, at the Wayback Machine.
10) Tulving, E. 1968 Theoretical issues in free recall. In T. R. Dixon & D. L. Horton (Eds.). Verbal behavior and general behavior theory. Englewood Cliffs, N. J.: Prentice-Hall.
11) Tulving, E. 1972 Episodic and semantic memory. In E. Tulving & W. Donaldson (Eds.). Organization of memory. New York:Academic Press.
12) Tulving, E. 1974 Recall and recognition of semantically encoded words. Journal of Experimental Psychology, 102, 778-787.
13) Tulving, E. 1976a Ecphoric processes on recall and recognition. In J. Brown (Ed.). Recall and recognition. London: Wiley.
14) Papadopoulos, Dim P., et al. "Training object class detectors with click supervision." Proceedings of the IEEE Conference on Computer Vision and Pattern Recognition. 2017.
15) Jackson, Frank. (1982) "Epiphenomenal Qualia", Philosophical Quarterly, vol. 32, pp. 127-36.
16) Bill Faw. Consciousness, modern scientific study of. In Tim Bayne, Axel Cleeremans, and Patrick Wilken, editors, The Oxford companion to consciousness. Oxford University Press, 2014.
17) Dehaene S, Changeux JP. Reward-dependent learning in neuronal networks for planning and decision making. Prog Brain Res. 2000;126:217-29.
18) Masafumi Oizumi, Larissa Albantakis, and Giulio Tononi. From the phenomenology to the mechanisms of consciousness: Integrated Information Theory 3.0. PLOS Computational Biology, 10(5):1–25, 2014.
19) Sin-ichi Inage, Hana Hebishima, Application of Monte Carlo stochastic optimization (MOST) to deep learning, Mathematics and Computers in Simulation 199 (2022) 257–271, 2022.
20) Tononi G. "An Information Integration Theory of Consciousness". BMC Neuroscience, 5:42, 2004.
21) Leonid I. Perlovsky, Toward physics of the mind: Concepts, emotions, consciousness, and symbols, Physics of Life Reviews 3 (2006) 23–55.
22) Christopher Durugboa, Ashutosh Tiwari, Jeffrey R. Alcock, Modelling information flow for organisations: A review of approaches and future challenges, International Journal of Information Management, Volume 33, Issue 3, June 2013, Pages 597-610.
23) Hiroshi Ezawa, Physics Perspectives (in Japanese), BAIFUKAN CO., LTD, 1983. ISBN4-563-02160-1 C3042.

## Appendix－Derivation of the governing equation of decision making

The following summarizes the formulation by E. Nelson. The formulation was based on the description by H. Ezawa (E. Ezawa, 1983). First, the following Fokker-Planck equation is used.

$$\left[\frac{\partial}{\partial t} + \frac{\partial}{\partial x}b(x,t) - \frac{\nu}{2}\frac{\partial^2}{\partial x^2}\right]\rho(x_0, t_0|x, t) = 0 \qquad \text{A.1)}$$

Where, t is time, x is defined as Kullback–Leibler divergence. The second term on the left hand side has the following meanings:

$$\frac{\partial}{\partial x}[b(x,t)\rho] \qquad \text{A.2)}$$

Here, since no transition can occur without time, at t = $t_0$ the following is satisfied, which is the initial condition:
$$\rho(x_0, t_0 | x, t) = \delta(x - x_0) \quad \text{A.3)}$$
Moreover, $\rho$ satisfies the following conditions.
$$\int \rho(x_0, t_0 | x, t) \, dx = 1 \quad \text{A.4)}$$
In Eq. 5), the equation with time reversed is as follows:
$$\left[-\frac{\partial}{\partial t} + \frac{\partial}{\partial x} b^*(x,t) + \frac{v}{2}\frac{\partial^2}{\partial x^2}\right]\rho = 0 \quad \text{A.5)}$$
By summing A.1) and A.3), we obtain:
$$\frac{\partial}{\partial x}\left[-(b - b^*) + v\frac{\partial^2}{\partial x^2}\right]\rho = 0 \quad \text{A.6)}$$
This indicates that the values in [ ] are independent of x. Consider introducing the following u, v.
$$u = \frac{1}{2}(b - b^*), v = \frac{1}{2}(b + b^*) \quad \text{A.7)}$$
Using the findings from A.6) and A.7), we obtain the following:
$$u = \frac{v}{2}\frac{1}{\rho}\frac{\partial \rho}{\partial x} = \frac{v}{2}\frac{\partial}{\partial x}\ln(\rho) \quad \text{A.8)}$$
Further, replacing A.1) and A.5) with the expression of v yields:
$$\frac{\partial \rho}{\partial t} + \frac{\partial}{\partial x}(v\rho) = 0 \quad \text{A.9)}$$
Erasing $\rho$ using A.8) and A.9) yields the following equation for only u, v:
$$\frac{\partial u}{\partial t} = -\frac{v}{2}\frac{\partial^2 v}{\partial x^2} - \frac{\partial}{\partial x}(uv) \quad \text{A.10)}$$
Eq. 17) in the text and A. 10) are written together as follows:
$$\frac{\partial v}{\partial t} = \frac{v}{2}\frac{\partial^2 u}{\partial x^2} - v\frac{\partial u}{\partial x} + u\frac{\partial v}{\partial x} - \frac{1}{m}\frac{\partial V}{\partial x} \quad \text{A.11)}$$
$$\frac{\partial u}{\partial t} = -\frac{v}{2}\frac{\partial^2 v}{\partial x^2} - v\frac{\partial u}{\partial x} - u\frac{\partial v}{\partial x} \quad \text{A.12)}$$
We define the function $\chi$ below.
$$\chi(x,t) = u(x,t) + iv(x,t) \quad \text{A.13)}$$
where i is an imaginary unit. Using this $\chi$, A. 12) and A. 13) are unified as follows:
$$-\frac{\partial \chi}{\partial t} = \frac{v}{2}\frac{\partial^2 \chi}{\partial x^2} + \frac{1}{2}\frac{\partial^2 \chi}{\partial x^2} - \frac{1}{m}\frac{\partial V}{\partial x} \quad \text{A.14)}$$
In addition, the following transformations are performed.
$$\chi = \frac{v}{2}\frac{1}{\Psi}\frac{\partial \Psi}{\partial x} = \frac{v}{2}\frac{\partial}{\partial x}\ln(\Psi) \quad \text{A.15)}$$
The characteristics of this transformation are described below.
a) time derivative of equation A. 15):
$$-i\frac{\partial \chi}{\partial x} = -iv\frac{\partial}{\partial x}\frac{\partial}{\partial t}\ln(\Psi) = -iv\frac{\partial}{\partial x}\left(\frac{1}{\Psi}\frac{\partial \Psi}{\partial t}\right) \quad \text{A.16)}$$
b) spatial derivative of equation A. 15):
$$v\frac{\partial \chi}{\partial x} = -\frac{v^2}{\Psi^2}\left(\frac{\partial \Psi}{\partial x}\right)^2 + \frac{v^2}{\Psi}\frac{\partial^2 \Psi}{\partial x^2} \quad \text{A.17)}$$
Taking advantage of the fact that the first term on the right-hand side is $-\chi^2$, further differentiation yields:
$$\frac{v}{2}\frac{\partial^2 \psi}{\partial x^2} + \frac{1}{2}\frac{\partial \chi^2}{\partial x} = \frac{v^2}{2}\frac{\partial}{\partial x}\left[\frac{1}{\Psi}\frac{\partial^2 \Psi}{\partial x^2}\right] \quad \text{A.18)}$$

Substituting A. 16), A. 17) for A. 14) yields:

$$\frac{\partial}{\partial x}\left[iv\frac{1}{\Psi}\frac{\partial \Psi}{\partial t}+\frac{v^2}{2}\frac{1}{\Psi}\frac{\partial^2 \Psi}{\partial x^2}-\frac{1}{m}V\right]=0 \qquad \text{A.19)}$$

Since this expression means that the values in [ ] are independent of the space x, we put it equal to the time-only function, $\eta$ (t). Then we get:

$$iv\frac{\partial \Psi}{\partial t}=\left[-\frac{v^2}{2}\frac{\partial^2 \Psi}{\partial x^2}+\frac{1}{m}V+\eta\right]\Psi \qquad \text{A.20)}$$

Furthermore, we transform $\Psi$ using $\psi$ and $\eta$ as follows:

$$\Psi(x,t)=\psi(x,t)exp\left(-\frac{i}{v}\int \eta(s)ds\right) \qquad \text{A.21)}$$

In this transformation, $\eta$ (t) is removed and finally becomes an equation of only $\Psi$ below.

$$iv\frac{\partial \psi}{\partial t}=\left[-\frac{v^2}{2}\frac{\partial^2 \psi}{\partial x^2}+\frac{1}{m}V\right]\psi(x,t) \qquad \text{A.22)}$$

This becomes the governing equation that determines the ultimate "will."

$$\chi=v\frac{\partial}{\partial x}ln(\psi) \qquad \text{A.23)}$$

Next, the relation between the distribution function $\rho$ (x, t) and the function $\Psi$ (x, t) is obtained when the 'intention' settles into one. First, from 19) and 25), we obtain:

$$\frac{v}{2}\frac{\partial}{\partial x}ln(\rho)=\frac{1}{2}(\chi+\chi^*) \qquad \text{A.24)}$$

where $\chi^*$ is the complex conjugate of $\chi$. Therefore,

$$\frac{\partial}{\partial x}[ln(\rho)-ln[\psi]^2]=0 \qquad \text{A.25)}$$

$$ln(\rho)=ln[\psi]^2, \quad \therefore \rho(x,t)=[\psi(x,t)]^2 \qquad \text{A.26)}$$